\documentclass[runningheads]{llncs}
\usepackage{graphicx}
\usepackage[table,xcdraw]{xcolor}
\usepackage{hyperref}
\usepackage{booktabs}
\usepackage{amsmath}
\usepackage{amssymb}
\usepackage{multicol}
\usepackage{multirow}

\begin{document}

\title{ReFuSeg: Regularized Multi-Modal Fusion for Precise Brain Tumour Segmentation}
\titlerunning{ReFuSeg for Brain Tumour Segmentation}
\authorrunning{A. Kasliwal et al.}
\author{
  Aditya Kasliwal\textsuperscript{1},
  Sankarshanaa Sagaram\textsuperscript{1,**},
  Laven Srivastava\textsuperscript{2,**},
  Pratinav Seth\textsuperscript{1},
  and Adil Khan\textsuperscript{3}
}

\institute{
  Manipal Institute of Technology, \\
  Manipal Academy of Higher Education, Manipal, India \\
  \textsuperscript{1} Department of Data Science and Computer Applications \\
  \textsuperscript{2} Department of Computer Science \\
  \textsuperscript{3} Department of Mechanical Engineering
}

\maketitle

\begin{center}
\texttt{\{kasliwaladitya17,sanky.sagaram,lavensri,seth.pratinav,adilk5020\}@gmail.com}
\end{center}

\renewcommand{\thefootnote}{\fnsymbol{footnote}}
\footnotetext{**Authors have contributed equally.}

\begin{abstract}
Semantic segmentation of brain tumours is a fundamental task in medical image analysis that can help clinicians in diagnosing the patient and tracking the progression of any malignant entities. Accurate segmentation of brain lesions is essential for medical diagnosis and treatment planning. However, failure to acquire specific MRI imaging modalities can prevent applications from operating in critical situations, raising concerns about their reliability and overall trustworthiness.
This paper presents a novel multi-modal approach for brain lesion segmentation that leverages information from four distinct imaging modalities while being robust to real-world scenarios of missing modalities, such as T1, T1c, T2, and FLAIR MRI of brains. Our proposed method can help address the challenges posed by artifacts in medical imagery due to data acquisition errors (such as patient motion) or a reconstruction algorithm's inability to represent the anatomy while ensuring a trade-off in accuracy. Our proposed regularization module makes it robust to these scenarios and ensures the reliability of lesion segmentation.
\end{abstract}
\section{Introduction}
The rise of Artificial Intelligence in healthcare has made AI-based interventions for brain tumour diagnosis and pre-assessment increasingly vital. Analyzing brain tumours through AI-driven techniques contributes significantly and helps in understanding the progression of brain tumour cells and assisting in surgical groundwork. Characterization of these segmented tumours can directly aid in predicting the interim duration for diagnosis and the patient's overall life expectancy, making brain tumour segmentation crucial for various applications in this field.
Magnetic Resonance Imaging (MRI) is a reliable diagnostic tool that is crucial in monitoring and planning brain tumour surgeries. The recent advancements in automated brain tumour segmentation using MRI have achieved remarkable success and practical utility. \cite{myronenko} \cite{MRIarticle} These algorithms typically rely on multiple modalities, with the four most relevant being T1-weighted images with and without contrast enhancement, T2-weighted images, and FLAIR images. Combining these complementary 3D MRI modalities, such as T1, T1 with contrast agent (T1c), T2, and Fluid-attenuated Inversion Recovery (FLAIR), helps highlight different tissue properties and regions where the tumour has spread. The integration of multiple modalities is essential for capturing a comprehensive view of the brain and improving segmentation accuracy. Each modality provides unique insights into the underlying tissue properties and pathology, allowing the model to exploit complementary information for robust and precise lesion segmentation.

While deep learning-based brain tumour segmentation techniques have shown impressive performance in various benchmarks, they face challenges due to the limited kernel size in typical image segmentation models. \cite{myronenko} \cite{Konstantinos} This limitation hinders their ability to learn long-range dependencies necessary for accurately segmenting tumours of various shapes and sizes. In clinical routines, missing MRI sequences due to time constraints or image artifacts can be a common challenge. Therefore, developing methods that can compensate for missing modalities and recover segmentation performance is highly desirable, promoting the broader adoption of these algorithms in clinical practice.

We propose ReFuSeg: Regularized Multi-Modal Fusion for Precise Brain Lesion Segmentation, our proposed architecture utilizes a novel approach toward contrastive regularisation to learn features between multiple modalities. Our approach prevents the model from overfitting to any particular modality and promotes the learning of complementary information from each modality. \\ This leads to more robust and generalizable features capable of capturing the intrinsic characteristics of each modality.  All four encoders work independently, learning individual features amongst each modality which helps it maintain robustness within its predictions as it is not dependent on any one singular modality to make accurate segmentation predictions.

The effectiveness of our proposed approach has been validated through experimental evaluations on the BraTS 23 dataset \cite{bakas2017advancing,labella2023asnr,Menze2015TheMB}. Our method displays robustness in accurately segmenting brain lesions, even in the case of missing modalities where they exhibit outstanding Dice and Hausdorff-95 scores, even when provided with only limited portions of the original data. Furthermore, the proposed model is unaffected by the inclusion of noise artifacts, which are commonly found in everyday clinical usage. Owing to these promising results, we demonstrate the suitability of our approach for real-world scenarios. 

\section{Related Works}
In \textbf{Multimodal Image Segmentation} previous methodologies have explored various approaches to address the segmentation challenge. Some studies have incorporated 3D network convolutions which fused  the correlation representations via attention guided mechanisms \cite{zhou2020brain}. A distinct approach was taken by Havaei et al. \cite{havaei2017brain}, where they constructed a unified model through a self-supervised training pipeline for each channel. Instead of using volumetric data, they directly passed 2D slices into the model encoders. Predictions from multiple channels were combined by merging feature maps, and mean, and variance were computed to achieve the final segmentation.
In another study \cite{10.3389/fncom.2019.00056}, a cascaded network \cite{malmi2015cabs} was employed. In the first stage, the tumour was segmented, and subsequent stages focused on learning substructures for more detailed segmentation.

\textbf{Missing Modalities} A popular technique \cite{ganin2015unsupervised} involves utilizing an adversarial loss on intermediate feature maps from two domains, facilitating knowledge transfer between these domains. Similarly, in a different work \cite{manders2018adversarial}, a class-specific adversarial loss was employed on feature maps to transfer a learned network from a source domain to a target domain. Generative models \cite{Zhang2001SegmentationOB}  have also been used in the past to synthesize missing modalities. Furthermore, self-supervised learning techniques have been utilized \cite{10.1007/978-3-030-20351-1_32} by randomly dropping modalities during training and leveraging the learned combined feature maps. These feature maps are then adjusted to match any encoder distributions to compensate for the missing data.

\textbf{Contrastive learning} is a powerful mechanism for representation learning. The core principle involves training a neural network to map similar inputs closer together in a learned feature space while simultaneously pushing dissimilar inputs further apart.\cite{chen2020simple} Contrastive loss minimizes dissimilarity between positive examples based on the learned features. Contrastive learning enables the effective learning of shared and discriminative representations across these modalities by capitalizing on the inherent similarities and distinctions across different modalities. A combination of inter and intra-modal feature learning \cite{yuan2021multimodal}has also been used to capture similarities and minimize disagreement between each modality. 
In biomedical segmentation, \cite{10.1007/978-3-031-16443-9_12} a network was proposed where one encoder learns features from T1 modality while the secondary segmentation backbone consisting of Convolution blocks takes in the multi-modal images and minimizes contrastive loss. 
\section{Methodology}
\begin{figure}[pt]
    \centerline{\includegraphics[width=12cm]{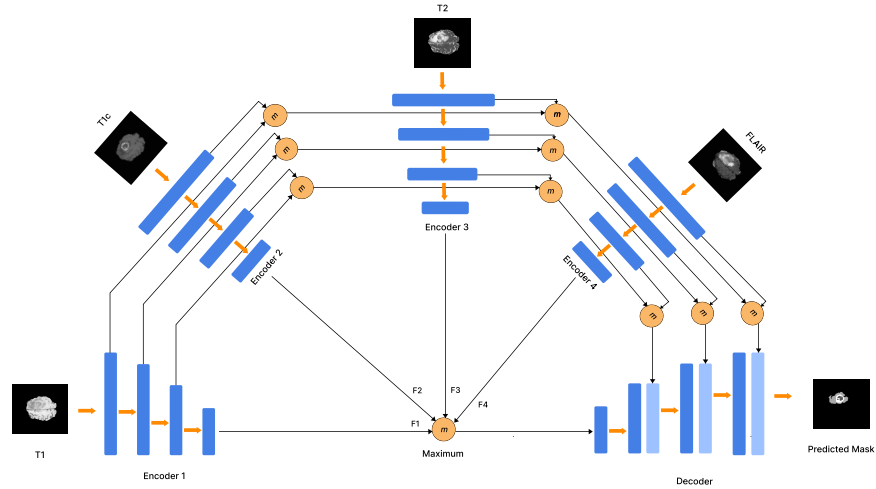}}
    \caption{Proposed ReFuSeg architecture for brain tumour segmentation.}
    \label{fig:ds2}
\end{figure}
In this paper, we introduce a novel methodology for brain lesion segmentation which is robust in the case of missing modalities due to the independent working of all the encoders, as a solution for the BraTS 2023 Adult Glioma Segmentation challenge \cite{bakas2017advancing,Menze2015TheMB,labella2023asnr} which involves generating segmented tumour masks by utilizing the four modalities provided in the dataset for each instance, i.e., T1, T1 with contrast agent (T1c), T2 and Fluid-attenuated Inversion Recovery (FLAIR). Our proposed approach, illustrated in the figures, comprises two primary components:
\begin{enumerate}
    \item  A multi-modal U-Net \cite{ronneberger2015u} based architecture incorporating four ResNet-34 \cite{koonce2021resnet} encoders alongside a feature fusion technique.
    \item A regularization module that receives final features from the four encoders, after which contrastive loss is calculated in the backward pass, enabling robustness in case of missing modality.
\end{enumerate}
In line with the conventional U-Net design, these encoders down-sample the input image and capture features at multiple levels of abstraction. It's worth mentioning that the four encoders work independently to extract distinctive features.

\subsection{Contrastive Regularization}
In our proposed approach, the features extracted from the encoders are directed to four distinct contrastive modules. Each contrastive module comprises a series of layers, including average pooling, fully connected, and batch normalization, followed by another fully connected layer and batch normalization. The resulting outputs from these contrastive modules are instrumental in calculating the contrastive loss similar to the approach used in \cite{Kasliwal2023CoReFusionCR}, which is then added to our final loss function.
Notably, the T1 and T1c modalities have a higher likelihood of containing similar features due to the nature of their MRI image acquisition methods, i.e., the feature changes noticeable in both modalities are primarily made in the post-processing of the MRI scans; the same is true for T2 and FLAIR modalities. However, the learning process is unaffected by redundant features because contrastive loss is a regularizing mechanism for these modalities.It promotes efficient selection of features and enhances individual contributions during the learning process.

\subsection{Feature Transfer from Encoders to Decoder}
Once the features are extracted from all four encoders, a new feature map is generated by taking the element-wise maximum of the corresponding level feature maps from each encoder. This ensures that the most relevant features from each encoder are selected, resulting in a new feature map that effectively captures pertinent information from all modalities.
The decoder, which comprises several decoding blocks, then receives the new feature map through skip connections. During the upsampling process, the decoder combines these upsampled feature maps with their corresponding counterparts from the encoders, utilizing the skip connections. However, in this modified U-Net model, the skip connections transfer the newly calculated feature map from the four encoders rather than individual feature maps from each encoder.
By amalgamating features extracted from the T1, T1c, T2, and FLAIR MRI images across multiple levels of abstraction, the adapted U-Net model demonstrates its capability to generate high-quality outputs.

\subsection{Handling Missing Modalities}

When faced with missing modalities, such as the absence of a T1-weighted scan in the input, our model exhibits robust performance due to the independent functioning of the four encoders. The model effectively utilizes the available encoders corresponding to the present modalities in such situations. The multilevel feature maps extracted from these encoders are directly passed to the decoder via the skip connections.
This approach ensures the model can still capture relevant features from the available input data and produce accurate outputs, even in missing modalities.
\subsection{Loss Function}
\textbf{Dice Loss}
The Dice loss \cite{DBLP:journals/corr/MilletariNA16} serves as a metric to assess the overlap between binary predicted and ground truth masks in image segmentation tasks. Its objective is to maximize the similarity between these masks. Therefore, minimizing the Dice loss during model training leads to improved accuracy of the segmentation model. Additionally, our experiments demonstrate that combining Dice loss with contrastive loss further enhances the segmentation performance.
\begin{equation}
\text{Dice Loss} = 1 - \frac{2 \sum_{i=1}^{n} y_i \hat{y}i}{\sum_{i=1}^{n} y_i^2 + \sum_{i=1}^{n} \hat{y}_i^2}
\end{equation}
where $y_i$ represents the ground truth value for the i-th sample, and $\hat{y}_i$ represents the corresponding predicted value from the model. The summation runs over all n samples in the dataset.

\textbf{Focal Loss}
The Focal loss \cite{DBLP:journals/corr/abs-1708-02002} introduces a modulating term to the cross-entropy \cite{DBLP:journals/corr/abs-1805-07836} loss, aiming to prioritize learning on challenging misclassified examples. This dynamic scaling of the cross entropy loss involves the scaling factor diminishing to zero as confidence in the correct class rises. In the context of image segmentation tasks, the Focal loss assumes a pivotal role in handling class imbalance and accentuating challenging samples, resulting in notable improvements in segmentation performance. The Focal loss formula is given as follows:

\begin{equation}
\text{Focal Loss} = -\frac{1}{n} \sum_{i=1}^{n} \left( \alpha (1 - \hat{y}_i)^{\gamma} y_i \log(\hat{y}_i) + (1 - \alpha) \hat{y}_i^{\gamma} (1 - y_i) \log(1 - \hat{y}_i) \right)
\end{equation}

In the formula, $y_i$ represents the ground truth value (0 or 1) for the i-th sample, and $\hat{y}_i$ represents the corresponding predicted value from the model. The summation runs over all n samples in the dataset.

The adjustable hyperparameters, $\alpha$ and $\gamma$, enable us to control the focusing effect and the rate at which the loss decreases for well-classified samples. Integrating the Focal loss with other appropriate loss functions, such as the contrastive loss mentioned in our experiments, can improve segmentation results.

\textbf{Contrastive Loss}
Contrastive loss \cite{khosla2020supervised} serves as a fundamental loss function utilized in machine learning to train models for similarity learning. Its purpose is to facilitate learning data representations, wherein similar data points are brought closer together in the representation space while dissimilar data points are pushed further apart. This enhances image fidelity and realism by reducing artifacts and noise in the output images.
In their work, \cite{chen2020simple} applied this contrastive loss to train their model by comparing pairs of images and calculating a similarity score between them. The contrastive loss formula is defined as follows:

\begin{equation}
l(v_i,v_j) = -\log\left(\frac{\exp(\text{sim}(v_i, v_j))}{\sum_{k \neq i} \exp(\text{sim}(v_i, v_k))}\right)
\end{equation}

Here, $v_i$ and $v_j$ represent the representations of two modalities of the same instance and $\text{sim}(·, ·)$ denotes the function computing the cosine similarity. The denominator in the formula represents the sum over all views $k$ in the batch, excluding $i$.
Subsequently, the contrastive loss is computed over the corresponding instances of two modalities in the batch of size $N$ and is then averaged as:
\begin{equation}
L_{C} = \frac{1}{2N} \sum_{i=1}^{N} \left[ l(v_{i\text{x}}, v_{i\text{y}}) + l(v_{i\text{y}}, v_{i\text{x}}) \right] 
\end{equation}

$\qquad \qquad \qquad \qquad \qquad \qquad \forall  \left( x, y \right) = \left[{(T1, T1c), (T2, FLAIR)}\right]$

The resulting $L_C$ contributes to the overall loss function, which is critical in guiding the model's training process for similarity learning and image representation enhancement.

\textbf{Final Loss}
The final loss employed in our paper, achieved after rigorous evaluation of different hyperparameter combinations, is represented as follows:

\begin{equation}
L_{Final} = 0.5 \cdot L_{Dice} + 0.5 \cdot L_{Focal} + \beta \cdot L_{C}
\end{equation}

Here, $\beta$ acts as a switch for contrastive loss in the overall loss function.

\section{Experimental Analysis}
\begin{table}[pt]
\centering
\caption{Validation Results for BraTS 2023 Dataset.}
\label{tab:table1}
\begin{tabular}{|c|c|c|c|c|c|c|}
\hline
\textbf{Contrastive} & \multicolumn{3}{c|}{\textbf{Dice}} & \multicolumn{3}{c|}{\textbf{Hausdorff}} \\
\textbf{Regularisation} & \textbf{ET} & \textbf{TC} & \textbf{WT} & \textbf{ET} & \textbf{TC} & \textbf{WT} \\
\hline
$\times$  & 0.792 & 0.828 & 0.909 & 22.9 & 14.07 & 7.34 \\
\hline
$\checkmark$ & 0.786 & 0.832 & 0.910 & 21.8 & 9.17  & 7.08 \\
\hline
\end{tabular}
\end{table}

The BraTS \cite{bakas2017advancing,Menze2015TheMB,labella2023asnr} 2023 dataset was used in our study, which consisted of 5,880 MRI scans from 1,470 patients with brain diffuse glioma. The BraTS mpMRI scans were provided in NIfTI format (.nii.gz) and included native (T1) and post-contrast T1-weighted (T1c), T2-weighted (T2), and T2 Fluid Attenuated Inversion Recovery (FLAIR) volumes. The training dataset had 1,251 instances, while the validation dataset had 219 instances. We submitted our predicted results on the validation dataset to the challenge website to assess our model's performance.
The original 3D files, each of size 240x240x155, were preprocessed into 155 2D slices of dimensionality 240x240. This preprocessing step addressed spatial invariance, reduced computational complexity, and enhanced anatomical interpretability. Additionally, we applied several augmentations to the data, including horizontal and vertical flips with probabilities of 0.5, rotation with a limit of 20 degrees, shift limit of 0.1, and probability of 0.5, random crop to a size of 224x224, and final resizing to 240x240.

We used softmax as the activation function in the output layer.The performance of the model was evaluated using the Dice score and Hausdorff-95 distance. The Dice score is a metric that measures the overlap between two sets, and a high Dice score indicates that the model has accurately captured the boundaries and shapes of the target structure. The Hausdorff-95 distance, on the other hand, measures the distance between the nearest points of two sets, and it is specifically used to assess the model's performance at boundary regions. To comprehensively evaluate the model, both these metrics were employed in our analysis.
Our model was trained using the Adam optimizer \cite{kingma2014adam} with a learning rate of $10^{-4}$.
During the initial experimentation, the proposed model was trained for 50 epochs without contrastive regularization. Subsequently, another training run was conducted with contrastive regularization. The comparison of results between non-contrastive and contrastive regularization can be found in Table \ref{tab:table1}.

To evaluate the model's robustness when faced with missing modalities, we performed inference four times, each time excluding one of the four available modalities. We conducted this inference process for both non-contrastive and contrastive approaches, and the detailed results are presented in Table \ref{Table 2}.
\section{Results}
The validation results, reveal noteworthy insights into the impact of contrastive regularization on the model's performance. These findings, as illustrated in Table 1, demonstrate a remarkable and favorable improvement in the model's performance when contrastive regularization is employed, in comparison to its performance without this technique.

\begin{table}[]
\centering
\caption{Comparing impact of missing modalities on BraTS 2023 validation set, with
and without contrastive regularization.
}
\label{Table 2}
\begin{tabular}{|l|l|lll|lll|}
\hline
\multicolumn{1}{|c|}{\multirow{2}{*}{\textbf{\begin{tabular}[c]{@{}c@{}}Modality   \\ Dropped\end{tabular}}}} &
  \multicolumn{1}{c|}{\multirow{2}{*}{\textbf{\begin{tabular}[c]{@{}c@{}}Contrastive  \\  Regularisation\end{tabular}}}} &
  \multicolumn{3}{c|}{\textbf{Dice}} &
  \multicolumn{3}{c|}{\textbf{Hausdorff-95}} \\ \cline{3-8} 
\multicolumn{1}{|c|}{} &
  \multicolumn{1}{c|}{} &
  \multicolumn{1}{l|}{\textbf{ET}} &
  \multicolumn{1}{l|}{\textbf{TC}} &
  \textbf{WT} &
  \multicolumn{1}{l|}{\textbf{ET}} &
  \multicolumn{1}{l|}{\textbf{TC}} &
  \textbf{WT} \\ \hline
T1 &
  $\times$ &
  \multicolumn{1}{l|}{0.753} &
  \multicolumn{1}{l|}{0.824} &
  0.901 &
  \multicolumn{1}{l|}{40.01} &
  \multicolumn{1}{l|}{17.76} &
  14.18 \\ \hline
T1 &
  $\checkmark$ &
  \multicolumn{1}{l|}{\textbf{0.769}} &
  \multicolumn{1}{l|}{\textbf{0.833}} &
  \textbf{0.908} &
  \multicolumn{1}{l|}{\textbf{31.04}} &
  \multicolumn{1}{l|}{\textbf{13.26}} &
  \textbf{9.59} \\ \hline
T1c &
  $\times$ &
  \multicolumn{1}{l|}{0.074} &
  \multicolumn{1}{l|}{0.322} &
  \textbf{0.878} &
  \multicolumn{1}{l|}{253.92} &
  \multicolumn{1}{l|}{45.08} &
  \textbf{11.47} \\ \hline
T1c &
  $\checkmark$ &
  \multicolumn{1}{l|}{\textbf{0.762}} &
  \multicolumn{1}{l|}{\textbf{0.803}} &
  0.874 &
  \multicolumn{1}{l|}{\textbf{38.91}} &
  \multicolumn{1}{l|}{\textbf{22.06}} &
  30.83 \\ \hline
T2 &
  $\times$ &
  \multicolumn{1}{l|}{0.057} &
  \multicolumn{1}{l|}{0.271} &
  0.884 &
  \multicolumn{1}{l|}{200.89} &
  \multicolumn{1}{l|}{65.63} &
  \textbf{7.17} \\ \hline
T2 &
  $\checkmark$ &
  \multicolumn{1}{l|}{\textbf{0.783}} &
  \multicolumn{1}{l|}{\textbf{0.819}} &
  \textbf{0.89} &
  \multicolumn{1}{l|}{\textbf{27.41}} &
  \multicolumn{1}{l|}{\textbf{17.5}} &
  16.22 \\ \hline
FLAIR &
  $\times$ &
  \multicolumn{1}{l|}{0.745} &
  \multicolumn{1}{l|}{0.703} &
  0.401 &
  \multicolumn{1}{l|}{38.00} &
  \multicolumn{1}{l|}{26.85} &
  27.96 \\ \hline
FLAIR &
  $\checkmark$ &
  \multicolumn{1}{l|}{\textbf{0.768}} &
  \multicolumn{1}{l|}{\textbf{0.786}} &
  \textbf{0.524} &
  \multicolumn{1}{l|}{\textbf{27.01}} &
  \multicolumn{1}{l|}{\textbf{23.56}} &
  \textbf{21.97} \\ \hline
\end{tabular}
\end{table}

As indicated in Table 1, the model exhibited signs of overfitting on specific modalities when contrastive regularization was not utilized. In such cases, the model excessively relied on a limited subset of the available modalities. As a result, the model's overall performance suffered significantly during validation, especially when confronted with missing modalities. Conversely, incorporating contrastive regularization yielded substantial enhancement in the model's performance in the presence of missing modalities, demonstrating reduced reliance on any single modality (T1, T1c, T2, or T2 flair) and, instead, effectively harnessing features from all available modalities. This regularization effect signifies the model's adept utilization of features across all modalities, effectively mitigating the overfitting challenges.

\section{Conclusion}
This research paper introduces an innovative framework that integrates data fusion and regularization techniques for semantic segmentation, utilizing four encoders within a U-Net-based architecture. The proposed framework serves as our response to the BraTS 2023 Adult Glioma Segmentation challenge \cite{bakas2017advancing,Menze2015TheMB,labella2023asnr}, held at the 9th MICCAI Workshop on Brain Lesions (BrainLes). It represents a straightforward and resource-efficient architecture compared to other models in this field.

Our architecture remarkably yields  outstanding results with Dice scores of 0.786, 0.832, and 0.910, as well as Hausdorff distances of 21.8, 9.17, and 7.08 for the enhancing tumour (ET), tumour core (TC), and whole tumour (WT) classes respectively, on the validation dataset with contrastive regularization. The model possesses the ability to be robust in handling missing data and maintaining its efficacy even when dealing with absent modalities, as demonstrated in Table \ref{Table 2}. Notably, we believe to be the first to present such a fusion model that effectively addresses missing modalities, rendering our architecture highly suitable for real-world scenarios with frequently occurring missing data. Looking ahead, we are eager to explore further experiments involving the scaling of contrastive regularization. Additionally, we plan to investigate the applicability of this approach in diverse domains and industries, broadening its potential impact.

\section{Acknowledgments}

\bibliographystyle{splncs04}
\bibliography{bibfile}
\end{document}